\documentclass[prl,superscriptaddress,twocolumn]{revtex4}
\usepackage{enumerate}
\usepackage{amsfonts,amssymb,amsmath}
\usepackage[]{graphics,graphicx,epsfig}
\usepackage{amsthm}
\usepackage[colorlinks=true,linkcolor=blue,citecolor=blue]{hyperref}
\usepackage{marvosym}

\usepackage{pdfpages}
\usepackage{epstopdf}
\usepackage{fdsymbol}
\usepackage{graphicx}
\usepackage{dcolumn}
\usepackage{natbib}
\usepackage{color}
\usepackage{xcolor}
\usepackage{multirow}
\usepackage{ulem}
\newcommand{\ket}[1]{|{#1}\rangle}
\newcommand{\bra}[1]{\langle{#1}|}

\begin{document}


\title{Rabi transport and the other finite-size effects in one-dimensional discrete-time topological quantum walk}

\author{Andrzej Grudka}
\affiliation{Institute of Spintronics and Quantum Information, Faculty of Physics, Adam Mickiewicz University, 61-614 Pozna\'n, Poland}

\author{Marcin Karczewski}
\affiliation{Institute of Spintronics and Quantum Information, Faculty of Physics, Adam Mickiewicz University, 61-614 Pozna\'n, Poland}

\author{Pawe{\l} Kurzy{\'n}ski}
\affiliation{Institute of Spintronics and Quantum Information, Faculty of Physics, Adam Mickiewicz University, 61-614 Pozna\'n, Poland}

\author{Tomasz P. Polak}
\affiliation{Institute of Spintronics and Quantum Information, Faculty of Physics, Adam Mickiewicz University, 61-614 Pozna\'n, Poland}

\author{Jan W{\'o}jcik}
\affiliation{Institute of Theoretical Physics and Astrophysics, University of Gdańsk, 80-308 Gda\'nsk, Poland}
\affiliation{Institute of Spintronics and Quantum Information, Faculty of Physics, Adam Mickiewicz University, 61-614 Pozna\'n, Poland}

\author{Antoni W{\'o}jcik}
\affiliation{Institute of Spintronics and Quantum Information, Faculty of Physics, Adam Mickiewicz University, 61-614 Pozna\'n, Poland}

\date{\today}


\begin{abstract}
This paper investigates Rabi transport and finite-size effects in one-dimensional discrete-time topological quantum walks. We demonstrate the emergence of localized states at boundaries between topologically distinct phases and analyze how finite system sizes influence quantum walk dynamics. For finite lattices, we show that topology induces localized and bilocalized states, leading to Rabi-like transport as a result of degeneracy breaking due to finite-size effects. The study bridges the gap between topological protection and size-dependent dynamics, revealing transitions from ballistic motion to localized or oscillatory behavior based on the system's topological properties. Analytical and numerical methods are employed to explore the spectra and dynamics of quantum walks, highlighting the robustness of Rabi transport against disorder. The findings provide insights into controlled quantum transport and potential applications in quantum information processing.
\end{abstract}

\maketitle

\section{I Introduction}
Transport phenomena in quantum systems, particularly in one-dimensional chains such as spin chains, have gained significant attention due to their potential applications in quantum information processing, among many other prospective areas \cite{Bose2003,Bose2007,MULKEN2011}. Quantum chains can be realized in e.g. linear chains of semiconductor quantum dots \cite{Kandel2019,Kandel2021,Kandel2021v2,Burkard2023}, multielectron quantum dots \cite{Malinowski2019}, chains of superconducting transmon qubits \cite{Li2018,Wilen2021}, photonic lattices \cite{Perez2013,Chapman2016,Zhong2019,Smirnova2020,Chang2018}, Rydberg atom chains \cite{Leseleuc2019}, atoms trapped within photonic crystal waveguides \cite{Hung2016}, nuclear spin chains \cite{Zhang2005}.

A possible method of controlling and optimizing transport in these systems lies in the engineering of couplings between the effective spins. For example, by carefully designing the interaction strengths, one can obtain perfect state transfer \cite{Christandl2004,Duan2003,SERRA2022}. On the other hand, it has been shown \cite{Wojcik2005,Wojcik2007,Paganelli2006,Campos2007,Yao2011,Oh2010,Oh2011,Ji2024} that to obtain the perfect transfer one does not has to use engineered couplings and instead take the uniform couplings and utilize the emergence of the weak couplings at the ends of the chain. In this case, a pair of states emerges that are coupled to each other but effectively decoupled from the rest of the system.  Thus, coherent oscillations between states arise. This kind of dynamics (induced long-range effective interaction) is often called superexchange \cite{Yuan2024,Burkard2023,Oh2013,Lewis2023,Chan2023}. Moreover, these oscillations are profoundly robust against possible random disorder in the coupling parameters. Note that both methods had been executed experimentally, engineered coupling \cite{Zhang2005,Li2018} and weak coupling \cite{Qiao2021,Chan2021}.

Yet another promising method of obtaining the pair of decoupled states arose due to the topological properties of the quantum chains. The significance of topological thinking in physics was first recognized in the study of electronic properties within solid-state theory \cite{Kosterlitz:2017vt,Haldane:2017tt,Hasan:2010vc}. Since then, topological phenomena have been identified in a wide range of systems, including optical lattices \cite{Cooper:2019tr,Goldman:2016tw,Zhang:2018uz}, photonic structures \cite{Ozawa:2019us,Khanikaev:2013vy,Khanikaev:2017ut,Lu:2014us,Smirnova:2020ty}, and mechanical systems \cite{Ma:2019tr}. A common foundation for observing topological effects lies in the presence of translation symmetry and the resulting Bloch band structure.

In the context of quantum transport in one-dimensional systems, topological properties can be exploited to allow the formation of edge states at the ends of the quantum chain. Since these edge states are decoupled from the rest of the chain, one expects the formation of coherent oscillations. This phenomenon was investigated both theoretically \cite{Lang2017,Longhi2018,Almeida2016,Mei2018,Bello2016,Estarellas2017} and experimentally \cite{Kim2021}.

Our aim is to consider topologically protected coherent oscillatory motion from the perspective of discrete-time quantum walks (DTQWs) (see for review \cite{Kempe:2003aa}). Discrete-time quantum walks (DTQWs) have emerged as a powerful framework for studying quantum dynamics and simulating complex physical systems. A particularly intriguing aspect of DTQWs is their ability to host topological phases, which arise from the underlying structure of the walker's Hilbert space and the periodic evolution operator \cite{Harper:2020to,Kitagawa:2012wa,Kitagawa:2010vm}. The topological properties of DTQWs have been extensively studied theoretically \cite{Kitagawa:2010vm,Lam:2016wa,Asboth:2014up,Obuse:2011vf,Asboth:2012ub,Obuse:2015uv,Asboth:2013aa,Chen:2017tq,Moulieras:2013vc,Peng:2021wk,Rakovszky:2015um,Tarasinski:2014tv,Barkhofen:2017uu,Cardano:2016aa,Cardano:2017aa,Cedzich:2018vu,Cedzich:2021tf,Cedzich:2022to,Cedzich2018completehomotopy,Wang:2019tx,Ramasesh:2017uo,Asboth:2015we,Edge:2015ww,Mochizuki:2020wr,Panahiyan:2021ui,Khazali2022discretetimequantum,PhysRevA.107.032201} and experimentally \cite{Flurin:2017vp,Kitagawa:2012ur,Nitsche:2019vc,Xie:2020wu,Xu:2020wr,PhysRevLett.120.260501}, further highlighting their rich and versatile nature. Note that DTQWs have been experimentally realized on various platforms, such as ion traps \cite{Schmitz:2009wr,Zahringer:2010te}, superconducting systems \cite{Flurin:2017vp,Yan:2019ve}, nuclear magnetic resonance setups \cite{Du:2003vg,Ryan:2005wp}, optical lattices \cite{Karski:2009ty,Dadras:2018vo}, and linear optics \cite{Do:2005wa,Cardano:wm,Perets:2008tt,Tang:ub}. 

In this paper we focus on one-dimensional DTQW. We show the existence of localized states at the boundary between topologically distinct quantum walks. Moreover, we show how finite size effects affect the behavior of the quantum walk. We study both DTQWs for finite and infinite lattices, analytically and numerically, providing an extensive description on their dynamics and spectra. In the case of a finite lattice, we show how the topology of the system induces the existence of localized and bilocalized states. We show that Rabi-like transport arises as a consequence of degeneracy breaking due to finite-size effects. We highlight the rich crossover between topological protection and size-dependent dynamics, showcasing the transition between the typical ballistic motion of quantum walk to the localizations or Rabi oscillations dependent on the topological properties of the system.

This paper is organized as follows: Section II introduces the model of DTQW. The phenomena of localized states at the sharp boundary between topologically distinct phases are presented in section III, with finite size effects investigated in section IV. Section V is devoted to the analysis of ballistic and Rabi transport in DTQW. Section VI summarizes the article's findings. Lastly, in the appendix, we show rigorous analytical solutions of the DTQW on a finite chain.

\section{II Model}
\begin{figure}[b]
\includegraphics[width=8cm]{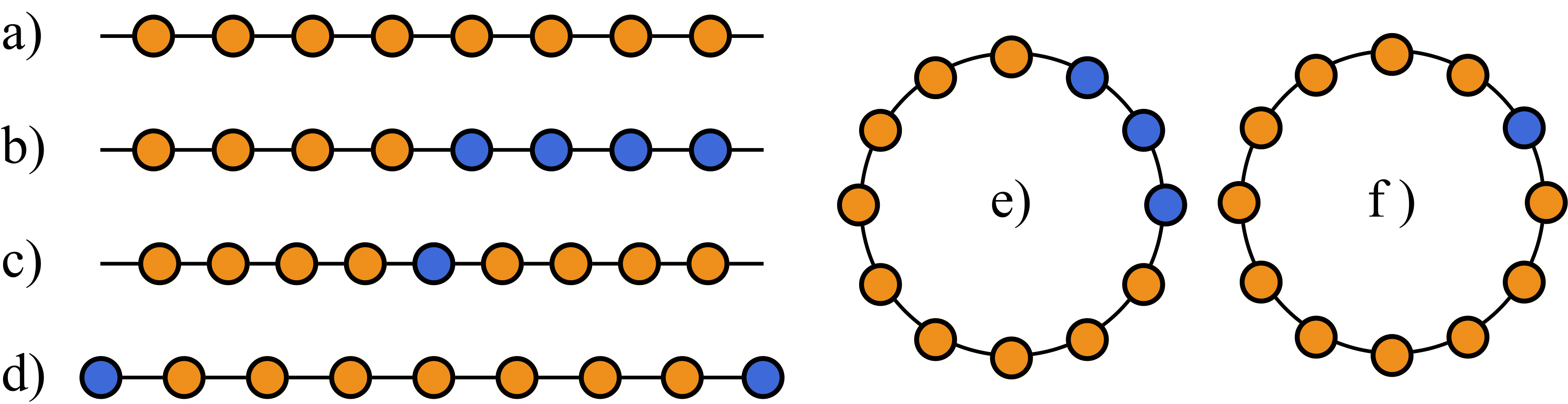}
\caption{Sketches of the models. Colors of the vertices indicate differences in the coin operators. (a) Homogeneous infinite chain. (b) Two homogeneous chains glued together. (c) Infinite chain with a single site defect. (d) A finite chain with reflective boundary coins. (e) A finite ring consisting of two homogeneous chains glued together. (f) Finite ring with a single site defect.}
\label{fig1}
\end{figure}

In this paper, we investigate the DTQW, which simulates the dynamics of a particle moving on a lattice. In this model, the state of the particle has two degrees of freedom. Thus, we describe the particle by its position and the internal two-dimensional "coin" state. The state vector in this basis has the form
\begin{equation}
    \ket{\Psi(t)} = \sum_x \ket{x}\otimes\begin{pmatrix} a_x(t)\\b_x(t)\end{pmatrix}.
\end{equation}
In the DTQW the evolution is discrete, governed by the unitary evolution operator. One step of this evolution is given by
\begin{equation}
    \ket{\Psi(t+1)} = U \ket{\Psi(t)}.
\end{equation}
The evolution operator can be broken down into two unitary operations
\begin{equation}
    U = S\ C,
\end{equation}
namely the step operator $S$ and the coin-toss operator $C$. In principle, the step operator governs the movement of the particle. It has the form
\begin{equation}
    S = \sum_x \left(|x+1\rangle\langle x| \otimes \begin{pmatrix}1&0\\0&0\end{pmatrix} + |x-1\rangle\langle x| \otimes \begin{pmatrix}0&0\\0&1\end{pmatrix} \right),
\end{equation}
or using the notation $|+1\rangle = \begin{pmatrix}1\\0\end{pmatrix}$ and $|-1\rangle = \begin{pmatrix}0\\1\end{pmatrix}$,
\begin{equation}
    S|x,\pm 1\rangle = |x\pm 1,\pm 1\rangle.
\end{equation}
The coin-toss operator, on the other hand, leaves the position of the particle as it is, but changes the internal coin state. This operation can be in the position-dependent form
\begin{equation}
    C = \sum_x |x\rangle\langle x| \otimes C_x,
\end{equation}
where $C_x$ is an operator from $\mathcal{U}(2)$ group
\begin{equation}
    C_x = e^{-i\delta_x} \begin{pmatrix}e^{i \zeta_x}\cos\theta_x & e^{i (\zeta_x+\sigma_x)}\sin\theta_x\\-e^{-i (\zeta_x+\sigma_x)}\sin\theta_x & e^{-i \zeta_x}\cos\theta_x\end{pmatrix},
\end{equation}
that unitarily transforms the coin state of the particle. The coin state is usually taken as the position-independent rotation, e.g. $(\delta_x = 0,\zeta_x = 0,\sigma_x = 0,\theta_x = \theta)$
\begin{equation}
    C_x = e^{-i\theta \sigma_y} = \begin{pmatrix}\cos\theta&&\sin\theta\\-\sin\theta&&\ \cos\theta    \end{pmatrix}
\end{equation}
Thus, the step operator moves the particles to the left or the right, depending on the coin state, whereas the coin-toss operator changes the coin state. 
To study the properties of this system, we need to examine its spectrum. To this end, we shall use the notion of effective Hamiltonian defined by 
\begin{equation}
    U = e^{-iH_{\text{eff}}}.
\end{equation}
Note that the effective Hamiltonian eigenenergies $\omega$ are defined modulo $2\pi$. We opt to confine them within the so-called first Floquet zone ($-\pi\leq \omega \leq \pi$). To define topological properties, we consider translationally invariant systems (Fig. \ref{fig1} a),  with $\delta_x = \delta, \ \zeta_x = \zeta, \ \sigma_x = \sigma, \  \theta_x=\theta$. In this case, it can be shown that the system is gapless only for $\theta = 0, \ \pi$. As was proved \cite{PhysRevA.107.032201}, for each fixed $(\delta,\zeta,\sigma)$ triple, there are two distinct topological phases (well defined in the gapped case). These two phases can be distinguished by the sign of $\theta$.

\section{III Localized states at a single sharp interface}
While topological properties are rigorously defined for translationally invariant systems, their application extends to systems with interfaces or boundaries. Notably, topological reasoning enables us to anticipate the presence of edge states localized at the interface between two topologically distinct phases.

\begin{figure}[t]
\centering
\includegraphics[width=0.4\textwidth]{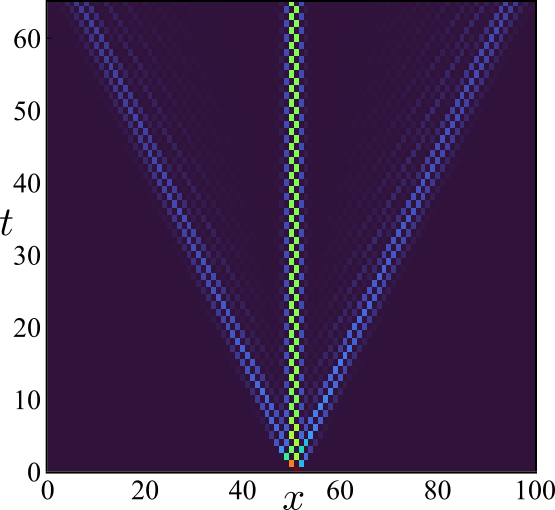}
\caption{Evolution of the inhomogeneous quantum walk $(\theta_\pm=\pm \pi/4, \  \delta=0, \ \zeta=0, \ \sigma=\pi/6)$ starting at the boundary ($x=0$).   The initial state is $\ket{\Psi(t=0)} =\frac{1}{\sqrt{2}}  \ \ket{0}\otimes\begin{pmatrix} 1\\i\end{pmatrix}$.}
\label{fig2}
\end{figure}
Let us now present the edge states of an inhomogeneous walk with a sharp boundary formed by two homogeneous walks glued together (Fig.\ref{fig1} b). To be specific for $x< 0$, $\theta_x = \theta_-$ whereas for $ x\geq0$, $\theta_x = \theta_+$. The explicit form of localized states in such a model was derived in \cite{PhysRevA.107.032201}. Note also that for a real, $\mathcal{SU}(2)$ coin $(\delta=0,\zeta=0,\sigma=0)$ with the use of a specific ansatz, localized states were also found in \cite{PhysRevA.93.052319}. For the general $\mathcal{U}(2)$ coin two localized states ($\eta=0, \pi$) can be written (\cite{PhysRevA.107.032201}) as

\begin{equation}\label{psin}
    \ket{\Psi_\eta} =\frac{1}{\sqrt{N}}(\ket{\Psi_\eta^+}+\ket{\Psi_\eta^-}),
\end{equation}
where unnormalized states $ \ket{\Psi_\eta^\pm}$ are given by
\begin{equation}\nonumber
    \ket{\Psi_\eta^+} = \sum_{x\geq 0} e^{i (\zeta+\eta) x} e^{-x/ \xi_+}\ket{x}\otimes \begin{pmatrix} 1 \\-e^{-1/ \xi_+}e^{-i\sigma x}\end{pmatrix},
\end{equation}

\begin{equation}
    \ket{\Psi_\eta^-} = \sum_{x < 0} e^{i (\zeta+\eta) x} e^{-x/ \xi_-}\ket{x}\otimes \begin{pmatrix} 1 \\-e^{-1/ \xi_-}e^{-i\sigma x}\end{pmatrix}.
\end{equation}
Normalization constant $N$ is given by 
\begin{equation}
    N=\sin^{-1} \theta_+-\sin^{-1} \theta_-,
\end{equation}

\noindent where, without losing generality, we assume that $\theta_-<0$ and $\theta_+>0$. Localization lengths $\xi_\pm$ are given by the equation
\begin{equation}\label{ksi}
    \xi_{\pm} = \ln^{-1}{\frac{\cos \theta_{\pm}}{1-\sin \theta_{\pm}}} .
\end{equation}
Energies of localized states are 
\begin{equation}\label{en}
    \omega = \delta + \eta.
\end{equation}
In Fig. \ref{fig2} we present the evolution of the walk $(\theta_\pm=\pm \pi/4, \  \delta=0, \ \zeta=0, \ \sigma=\pi/6)$ starting at the boundary ($x=0$).   The initial state is
\begin{equation}
    \ket{\Psi(t=0)} =\frac{1}{\sqrt{2}}  \ \ket{0}\otimes\begin{pmatrix} 1\\i\end{pmatrix}.
\end{equation}
\begin{figure}[b]
\includegraphics[width=8cm]{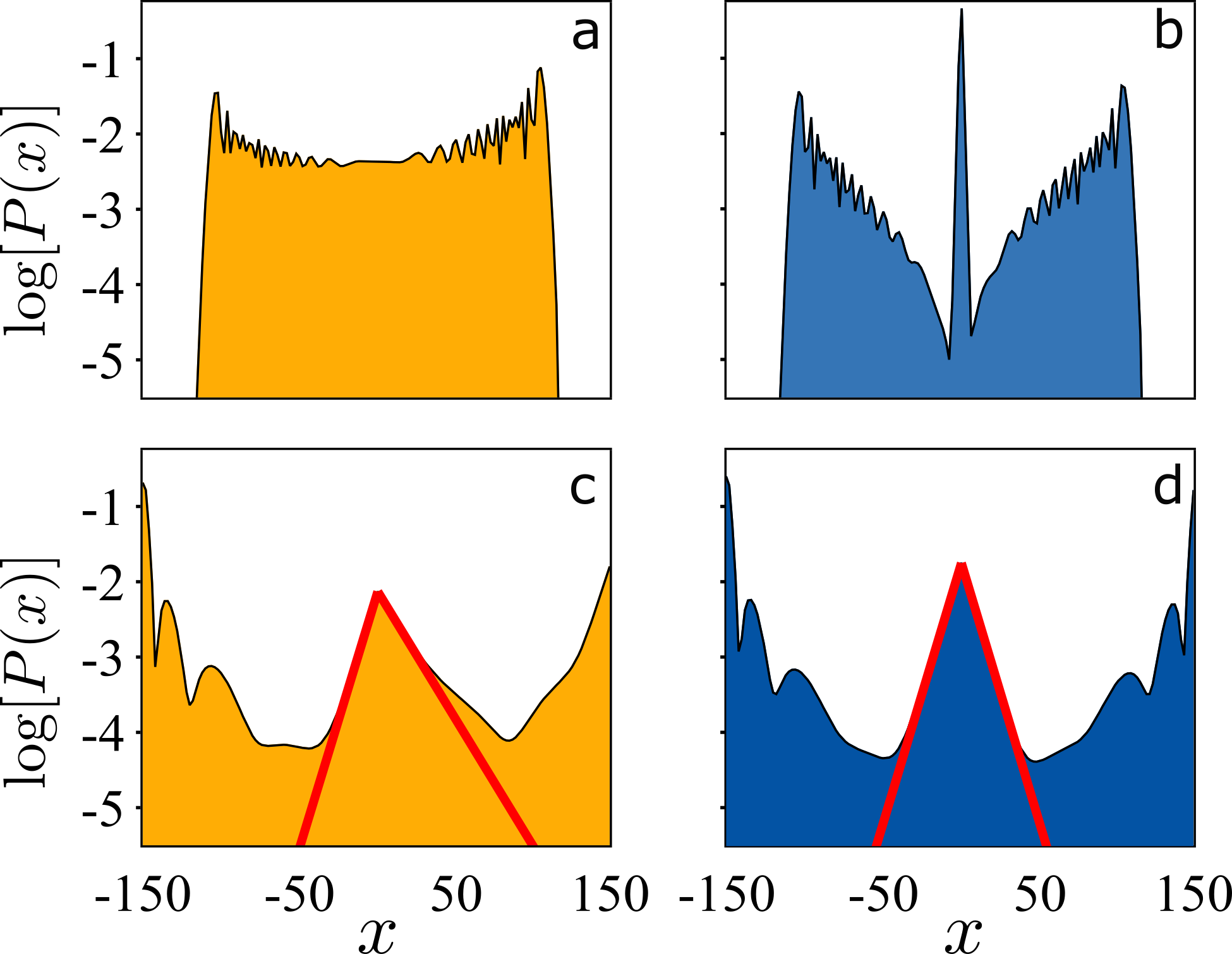}
\caption{Probability distribution for a homogeneous walk, with $\theta_{\mp} = \pi/4$, ($ \delta=0, \ \zeta=0, \ \sigma=\pi/6)$ (a) and an inhomogeneous ones, with $\theta_{\mp} = \mp \pi/4$ (b), $\theta_{-} = - \pi/40$, $\theta_{+} = \pi/80$ (c) and $\theta_{\mp} = \mp \pi/40$ (d). For the chosen time $t=150$, odd positions are not populated at all, so the probabilities of the only even positions are presented in the figure. Red lines (c,d) represent the approximation given by Eq. \ref{approx}. The initial state is $\ket{\Psi(t=0)} =\frac{1}{\sqrt{2}}  \ \ket{0}\otimes\begin{pmatrix} 1\\i\end{pmatrix}$.}
\label{fig3}
\end{figure}
One sees that part of the population is in ballistic motion, whereas the remaining part is trapped in the central region near the starting position. This trapped population oscillates with the period $T=2$. Note that for the initial state localized at $x=0$ 
\begin{equation}
c= \langle \Psi_{\eta=0}\ket{\Psi(t=0)}=\langle \Psi_{\eta=\pi}\ket{\Psi(t=0)} ,
\end{equation}
Where $\ket{\Psi_{\eta}}$ is defined in Eq. \ref{psin}. The initial state can be written as
\begin{equation}
    \ket{\Psi(t=0)} =c \ket{\Psi_{\eta=0}}+c \ket{\Psi_{\eta=\pi}}+\sqrt{1-2 |c|^2} \ket{\Psi_{band}},
\end{equation}
where $\ket{\Psi_{band}}$ is a state in the band (orthogonal to both localized states). $\ket{\Psi_{band}}$ gives rise to ballistic motion. Equally weighted superposition of two localized states with an energy difference $\Delta E=\pi$ gives rise to oscillations with the period $T=\frac{2 \pi}{\Delta E}=2$. The probability distribution of the walk for chosen time $t=150$ is presented on a logarithmic scale, in Fig. \ref{fig3} for both inhomogeneous and homogeneous quantum walks. In the case of an inhomogeneous walk, a localized population is visible in the center.
To underscore the sharp distinction between topological phases, we illustrate in Fig. \ref{fig3} (c,d) the probability distribution for an inhomogeneous quantum walk with parameters $\theta_+$ and $\theta_-$ differing only by a small margin.

\begin{figure*}[t]
\includegraphics[width=16cm]{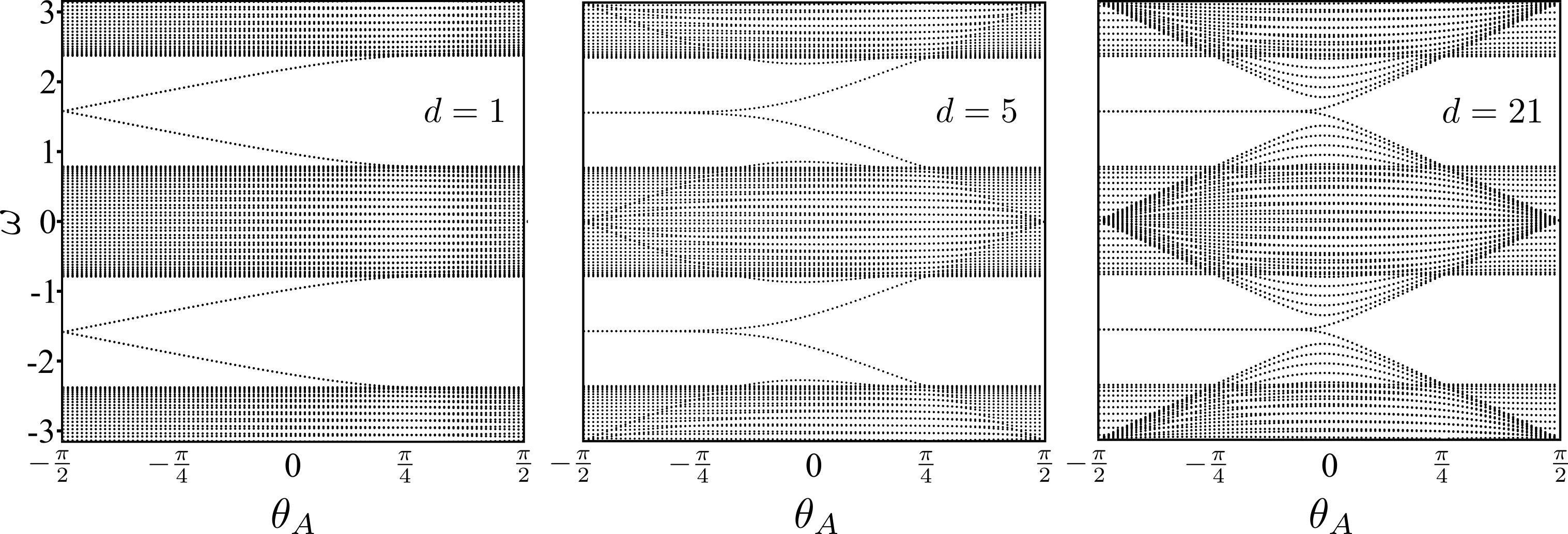}
\caption{Energies $\omega$ versus parameter $\theta_A$ for a walk on D-cycle ($D=42$) with two (topologically distinct) segments of different sizes $d$. (See Fig. \ref{fig1} (e,f)). Parameter $\theta_B$ is equal to $\pi/4$ whereas $\delta=-\pi/2,\zeta=-\pi/2,\sigma=0$.}
\label{fig7}
\end{figure*}

Moreover, in this figure, we show that the part of the probability distribution localized near the origin can be approximated by a simple formula
\begin{equation}\label{approx}
   p(x,t\gg1)= e^{-2|x|\theta_{\pm}}.
\end{equation}
That approximation is due to the fact that $\xi_{\pm}=\theta_{\pm}^{-1}$ in the limit of small $\theta_{\pm}$. 
The above is valid even in the asymmetric case $\theta_-\neq-\theta_+$ as seen in Fig. \ref{fig3} c. 

\section{IV Finite-size effects}

The sharp distinction between topological phases (as depicted in Fig. \ref{fig3} and rigorously established in \cite{PhysRevA.107.032201}) necessitates assuming an infinitely long lattice (referenced in Fig. \ref{fig1}b). Beyond a mere mathematical grasp of this limit, one must also comprehend the physical interpretation of such "infinity". From this perspective, segments of the lattice that are topologically distinct can be considered "infinite", provided their dimensions significantly exceed the localization length. Failure to meet this criterion can result in the blurring of boundaries between topological phases.

We illustrate the finite size effect in Fig. \ref{fig7}, which relate to walks on a cycle (Fig. \ref{fig1} (e,f)). This allows us to present energy values $\omega$ obtained through numerical diagonalization. The cycle comprises two topologically distinct segments, with lengths denoted as $d$ and $D-d$ respectively. The coin parameters $\theta_x$ for the first and second segments are $\theta_x=\theta_A$ and $\theta_x=\theta_B$ respectively. The remaining coin parameters are position-independent; for example, we select $(\delta=-\pi/2, \ \zeta=-\pi/2, \ \sigma=0)$.

\begin{figure}[b]
\includegraphics[width=8cm]{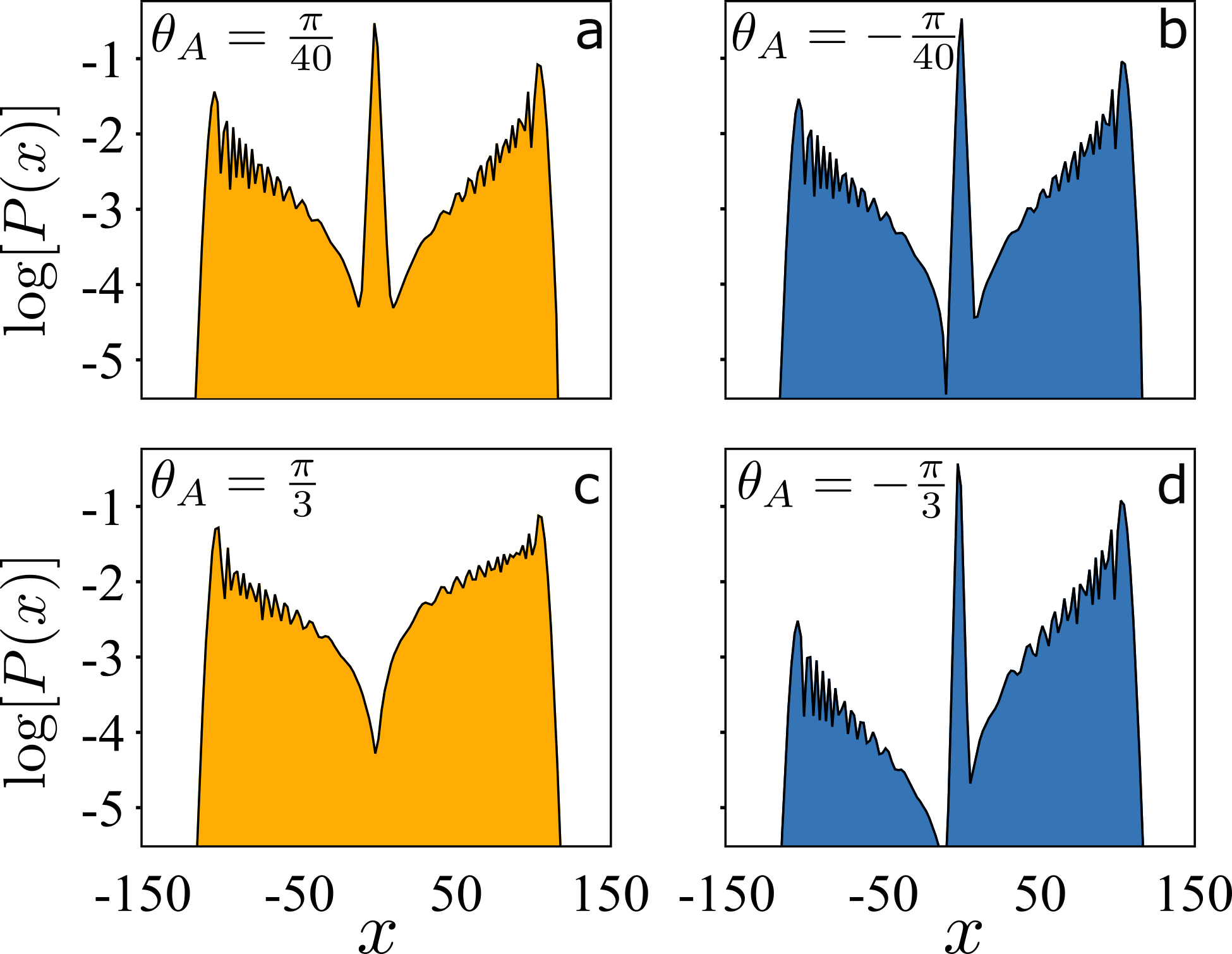}
\caption{Probability distribution of the quantum walk depicted in Fig. \ref{fig1}c with parameters $\theta_B=\pi/4$, with different $\theta_A$ (at position $x=0$). For the chosen time $t=150$, odd positions are not populated at all, so the probabilities of the only even positions are presented in the figure. The walk starts from the state $\ket{\Psi(t=0)} =  \frac{1}{\sqrt{2}}\ket{0}\otimes\begin{pmatrix} 1\\i\end{pmatrix}.$}
\label{fig8}
\end{figure}
The dependence of energy $\omega$ on parameter $\theta_A$ is depicted for fixed $\theta_B=\pi/4$. In the vicinity of $-\pi/2$, two degenerate pairs of states with energies $\omega=\pm \pi/2$ are observed (consistent with Eq. \ref{en}), reflecting the presence of two distinct boundaries. As $\theta_A$ increases, the degeneracy within each pair diminishes due to the finite size of the segments. Significantly, for a single defect ($d=1$), degeneracy breaks down at a much faster rate. Additionally, another finite-size effect arises: gap states persist even as 
$\theta_A$ surpasses zero and becomes positive. Hence, for small lattices, merely considering the sign of the parameters $\theta_A$ and $\theta_B$ is insufficient to ascertain the presence of gap states. The right part of Fig. \ref{fig7} demonstrates that for moderately sized lattices ($d=21$), the disappearance of gap states in the vicinity of $\theta_A=0$ is much more pronounced. 

The finite size effects observed in Fig. \ref{fig7} persist even in the limit of $D \to \infty$. As they result from the interplay between the segment and the localization length, not the ratio $d/D$ but rather $d/\xi_{\pm}$. To illustrate this, let us consider the case of an infinite chain with a single defect, i.e. $d=1$ (Fig. \ref{fig1}c). In Fig. \ref{fig8} we observe the existence of localized states for both positive and negative values of $\theta_A$. Interesting features become apparent for 
$\theta_A$ values that are far from $0$. Thus, we depict the distributions of the walk for $\theta_A= \pm \pi /3$. Here, we observe "anti-localization" (instead of the maximum, we have a minimum) when the defect is from the same topological phase as the rest of the line.

\section{V Ballistic and Rabi transport}
In this section, we consider the dynamics of a quantum walk on a finite lattice restricted by two totally reflecting coins (Fig \ref{fig1}d). Positions $x$ are integers in the range $x_{min},...,x_{max}$, thus $D=1+x_{max}-x_{min}$. At both ends of the lattice, the condition of total reflection is met with condition $\theta_{x_{min}}=\pm \pi/2$, $\theta_{x_{max}}=\pm \pi/2$.
\begin{equation}
    C(\theta =\pm \pi/2) = \pm \begin{pmatrix}0&1\\-1& 0    \end{pmatrix}.
\end{equation}

\begin{figure}[t]
\includegraphics[width=8cm]{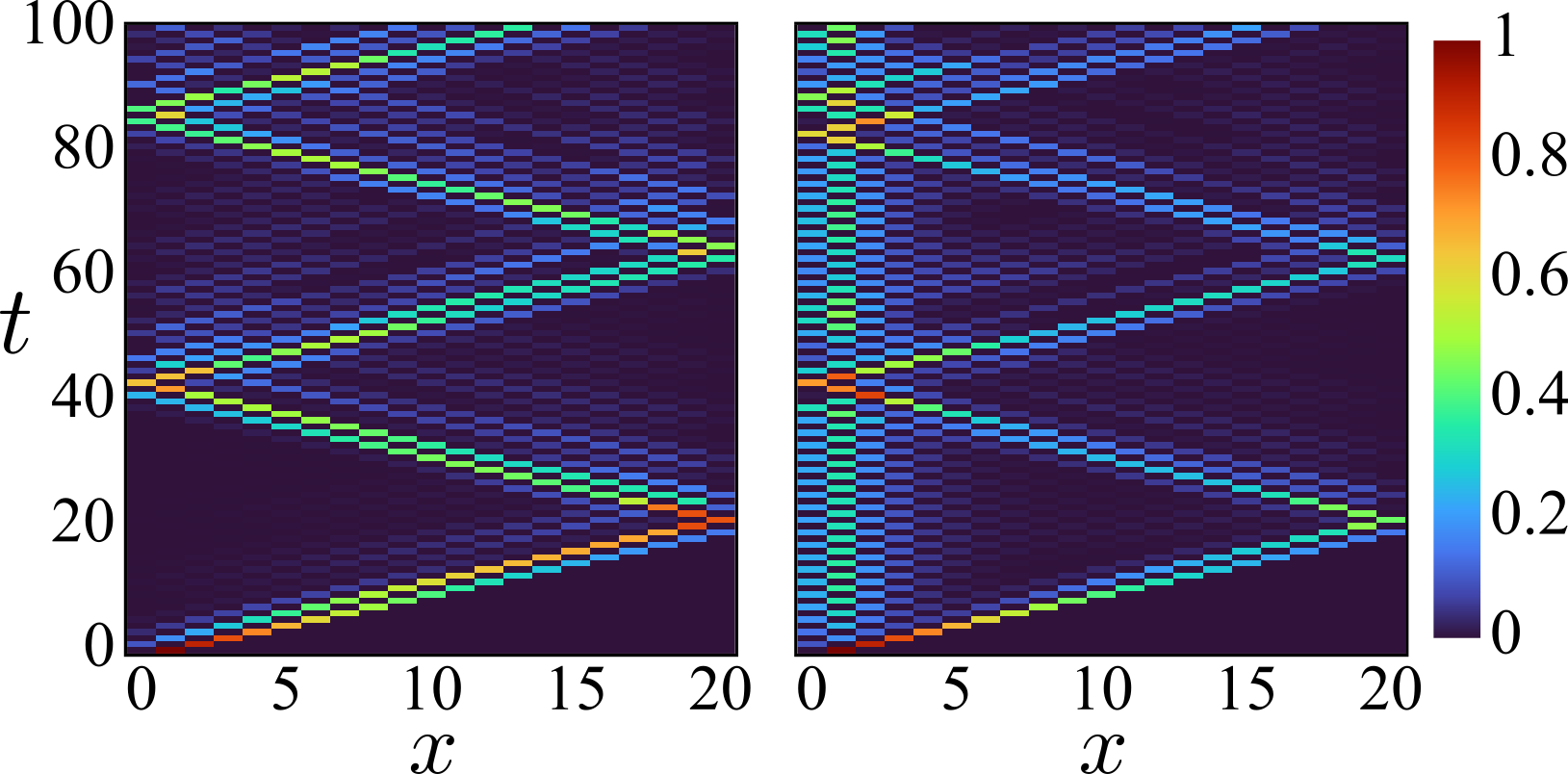}
\caption{Evolution of the quantum walk on the finite ($D=21$) wire (see Fig. \ref{fig1}d) with $\theta=\pi/10$, ($ \delta=0, \ \zeta=0, \ \sigma=0)$. On the left part  $\theta_{x_\text{min}}=\theta_{x_\text{max}}=\pi/2$ and on the right  $\theta_{x_\text{min}}=\theta_{x_\text{max}}=-\pi/2$. The initial state is $\ket{\Psi(t=0)} =  \ \ket{1}\otimes\begin{pmatrix} 1\\0\end{pmatrix}$.}
\label{fig10}
\end{figure}

"Bulk" coins, i.e. coins for $x \neq x_{min},x_{max}$ are characterized by position-independent parameters $\theta_x=\theta,\ \delta_x=  \delta, \ \zeta_x=  \zeta, \ \sigma_x=\sigma$.

Left part of Fig. \ref{fig10} presents the evolution of a walk ($\theta=\pi /10$) on a finite wire ($D=21$) with reflecting coins ($\theta_{min}=\theta_{max}=\pi/2$) that are in the same topological phase as the bulk. The initial state is 
\begin{equation}\label{in}
    \ket{\Psi(t=0)} =  \ket{x_{min}+1}\otimes\begin{pmatrix} 1\\0\end{pmatrix}.
\end{equation}
The evolution of this state results in the "ping-pong" dynamics, i.e. ballistic motion, bouncing between reflecting boundaries.

\begin{figure}[b]
\includegraphics[width=8cm]{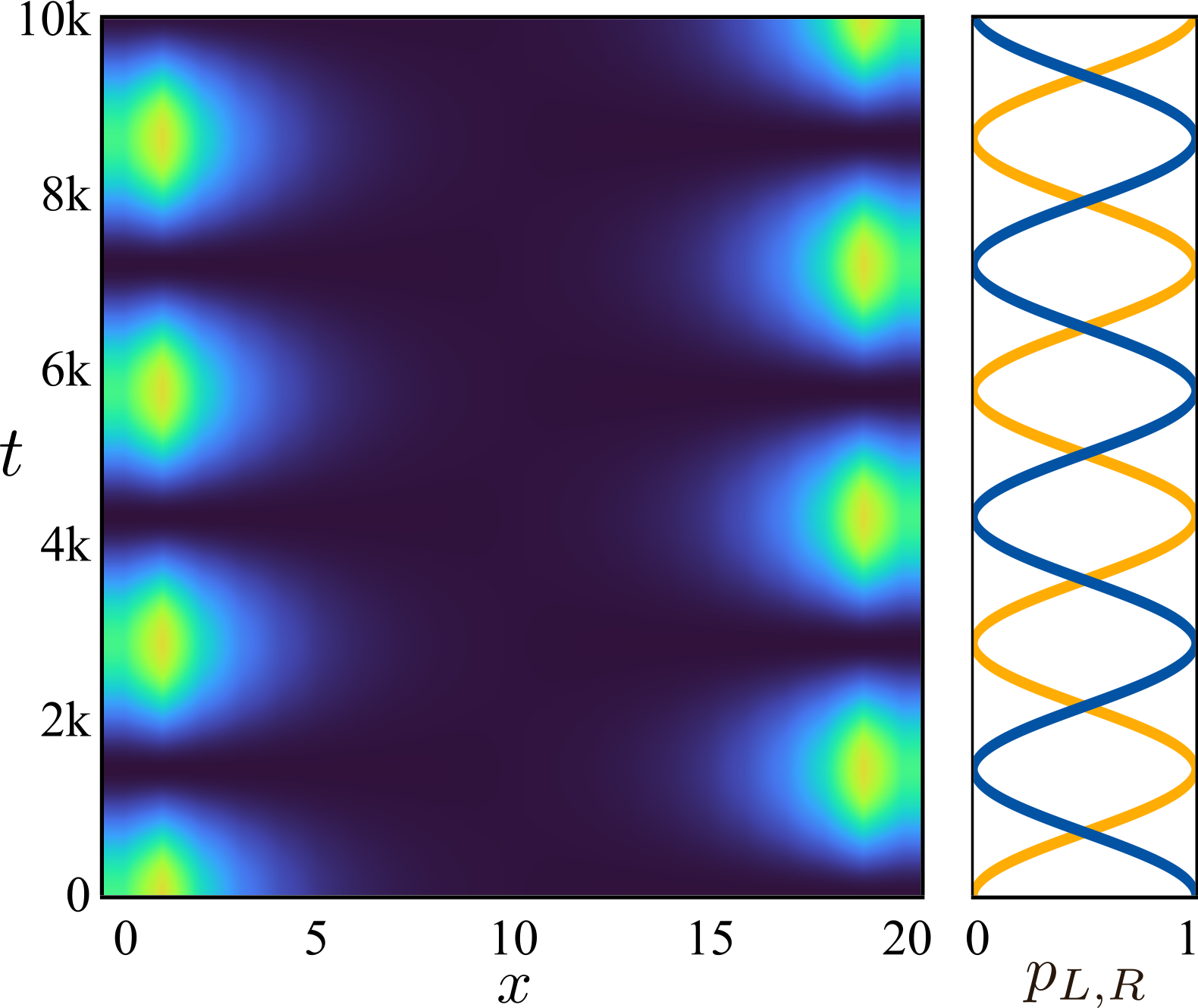}
\caption{Evolution of the quantum walk on the finite ($D=21$) wire with $\theta=\pi/10$, $\theta_{x_\text{min}}=\theta_{x_\text{max}}=-\pi/2$, ($ \delta=0, \ \zeta=0, \ \sigma=0)$ (left). Probability of populations of the states $\ket{\Psi_{L,R}}$ (right). The initial state is $\ket{\Psi(t=0)} =  \ket{\Psi_L}$.}
\label{fig12}
\end{figure}
If we change the sign of $\theta_{min}$ making the left boundary to be of topologically different phase from the bulk we observe in right part of Fig. \ref{fig10} aside from the "ping-pong" motion an additional feature, namely trapping a part of the population in an edge state in the vicinity of the left boundary. The initial state is again given by Eq. \ref{in}.

Quite a different situation is observed in Fig. \ref{fig12}. Here we observe the transport of the population from one end to another without populating the central part of the lattice. We call this phenomenon Rabi transport because, as we will show, it is based on coherent oscillations of a two-state superposition. Note that the time scale now is entirely different. 

Fig. \ref{fig12} represents a walk with both reflecting coins belonging to the topological phase different from the bulk ($\theta_{min}=\theta_{max}=-\pi/2$). The additional difference between the dynamics presented in Figs. \ref{fig10} and \ref{fig12} is the time scale and the initial state. Namely, this initial state for evolution presented in Fig. \ref{fig12}  is given by
\begin{equation}
    \ket{\Psi(t=0)} =\ket{\Psi_L},
\end{equation}
where $\ket{\Psi_L}$ is the following superposition of two eigenstates
\begin{equation}
   \ket{\Psi_L}=\frac{e^{-i \frac{\pi}{4}} \ket{\Psi_{\omega}}+e^{i \frac{\pi}{4}} \ket{\Psi_{-\omega}}}{\sqrt{2}}.
\end{equation}
Explicit forms of eigenstates $\ket{\Psi_{\pm \omega}}$ together with corresponding eigenvalues $\pm \omega$ are given in the Appendix.
Here, we just mention that these eigenvalues lie in the gap, in the vicinity of null energy, and eigenvectors are bi-localized at both ends of the lattice. 
\begin{figure*}[t]
\includegraphics[width=18cm]{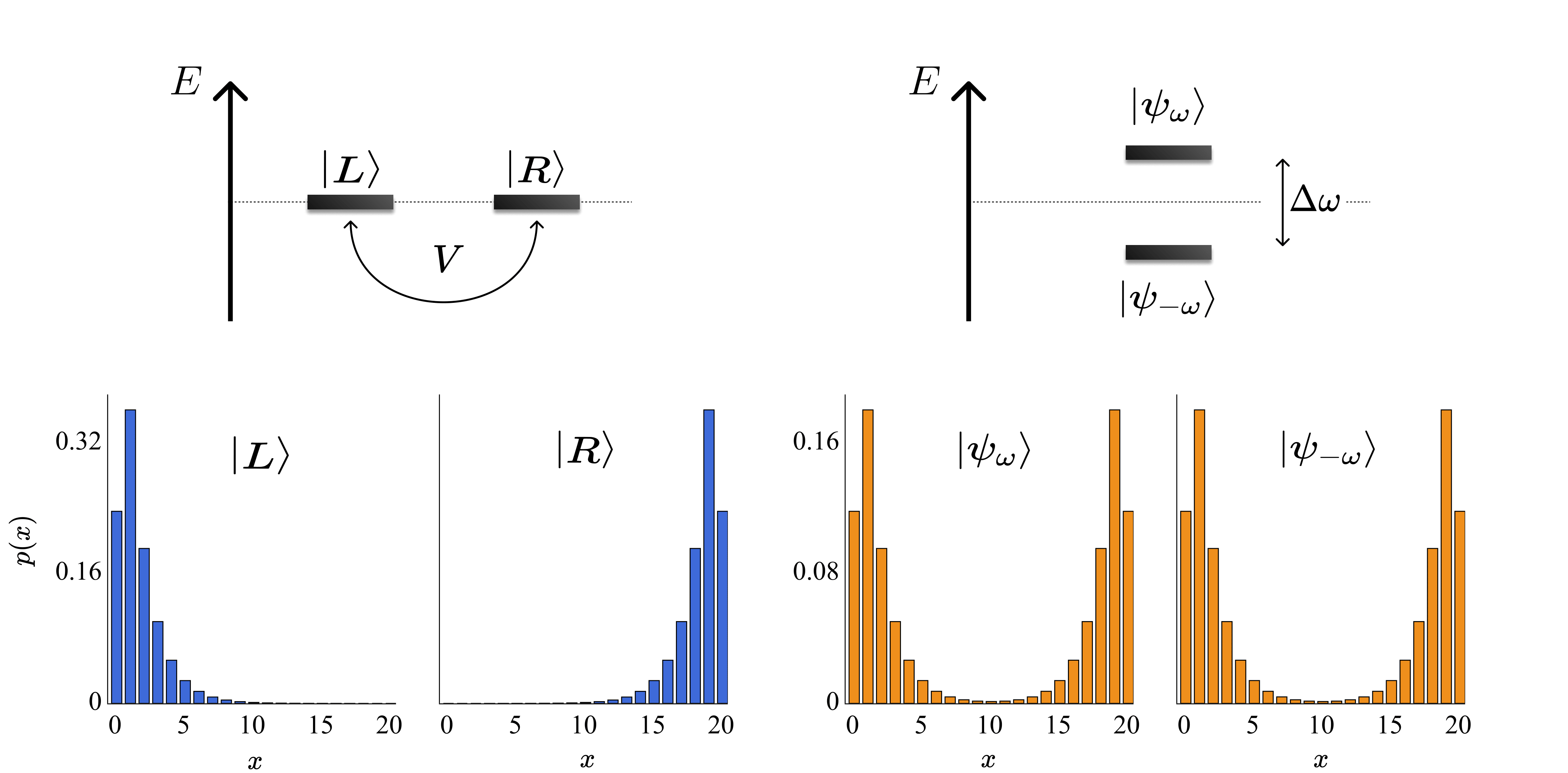}
\caption{Schematic view of coupled states $\ket{L}$ and $\ket{R}$, their energy levels and probability distribution on the wire (left). Eigenstates $\ket{\psi_\omega}$, $\ket{\psi_{-\omega}}$, their energy levels and probability distribution on the wire  (note the degeneracy breaking due to $V$ interaction) (right).}
\label{fig17}
\end{figure*}

Let us also define 
\begin{equation}
   \ket{\Psi_R}=\frac{e^{-i \frac{\pi}{4}} \ket{\Psi_{\omega}}-e^{i \frac{\pi}{4}} \ket{\Psi_{-\omega}}}{\sqrt{2}}.
\end{equation}

Note that $\ket{\Psi_R}$ and $\ket{\Psi_L}$ are mutually orthogonal. The distribution of population in these states, as well as in relevant eigenstates, is presented in Fig. \ref{fig17}.

 Probability of populations of the states $\ket{\Psi_{L,R}}$ during evolution is given by
\begin{equation}
  p_{L,R}(t)=|\langle \Psi (t)\ket{\Psi_{L,R}}|^2.
\end{equation}
These probabilities for dynamics presented in Fig. \ref{fig12} are shown on the right side of this figure. One sees ($p_{L}(t)+p_{R}(t)=1$) that dynamics is effectively confined to two-dimensional Hilbert space $H_{LR}$ spanned by $\ket{\Psi_{L, R}}$. This two-dimensional Hilbert space can be considered as effectively decoupled from the rest of the system. However, the remaining states form an effective coupling $V$ between states $\ket{\Psi_{L, R}}$. This coupling tends to zero (exponentially) when the length of the lattice increases. In the limit of infinite lattice states, $\ket{\Psi_{L, R}}$ forms a pair of degenerated eigenstates. In a finite-size case, on the other hand, the mentioned coupling removes degeneracy and the pair of eigenvectors $\ket{\Psi_{\pm \omega}}$ acquires non-zero energy difference $\Delta \omega=2 \omega$. A schematic view of a space $H_{LR}$ is presented in Fig. \ref{fig17}. Both states $\ket{\Psi_{L, R}}$ are equally weighted superpositions of two eigenvectors $\ket{\Psi_{\pm \omega}}$. Thus, if one of these states is initially populated, the population has to be exchanged between these states with period $T=\frac{2 \pi}{\Delta \omega}$. 
This difference is given by (for details, see Appendix)
\begin{equation}\label{appr}
\Delta \omega \equiv \frac{2 \sin 2 \theta}{(1+\sin \theta)^2}  \left(\frac{1+\sin \theta}{\cos \theta} \right)^{-2 D}
\end{equation}
and (as was mentioned) decreases exponentially with the size of the lattice $D$.

\begin{figure}[t]
\includegraphics[width=8cm]{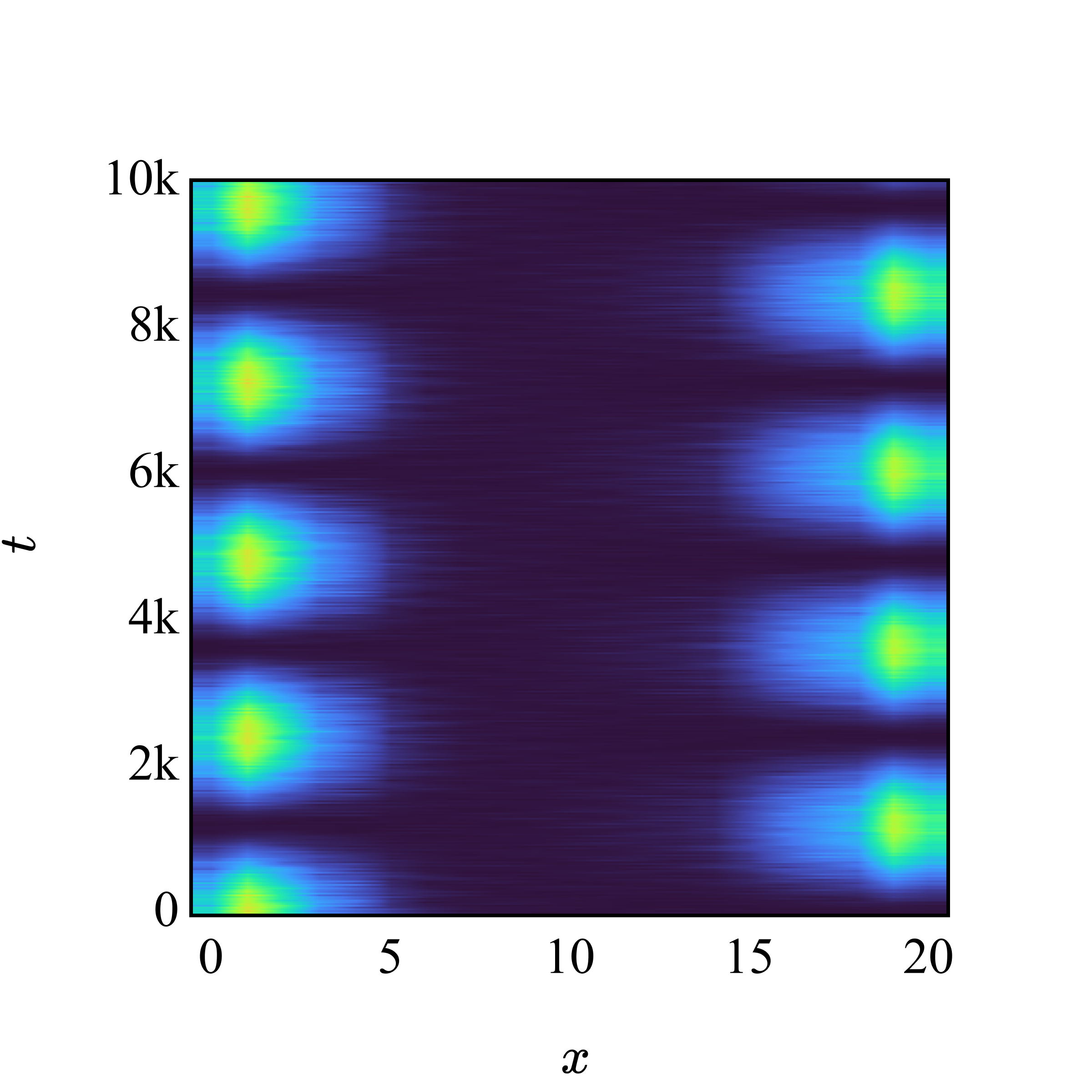}
\caption{Evolution of the noisy quantum walk on the finite ($D=21$) wire with uniformly distributed random $\theta\in(0,\frac{\pi}{5})$, $\theta_{x_\text{min}}=\theta_{x_\text{max}}=-\pi/2$, ($ \delta=0, \ \zeta=0, \ \sigma=0)$. The initial state is $\ket{\Psi(t=0)} =  \ \ket{1}\otimes\begin{pmatrix} 1\\0\end{pmatrix}$.}
\label{fig17a}
\end{figure}

Let us also emphasize that although eigenenergies can be calculated exactly (as is presented in the Appendix), the existence of gapped states is guaranteed just by topological arguments. So, one can expect that Rabi transport will be robust against randomness, which can be introduced in the model. To check this fact, we observe the dynamics of the walk similar to that presented previously, but this time with random values of $\theta_x$ in the bulk. However, we ensure that the values of all $\theta_x$ in the bulk are in the same topological phase (distinct from the phase of reflecting coins). We see in Fig. \ref{fig17a} that the main feature of Rabi transport is not affected by the introduced randomness. What changes is the period of oscillation as well as the total probability confined in the subspace $H_2$, because the previously defined initial state $\ket{\Psi_L}$ has now some nonzero overlap with band eigenstates.

\section{VI Summary}
We investigate the topological properties of discrete-time quantum walks (DTQWs) and their implications for quantum transport phenomena. By analyzing both the spectra and eigenstates of the system's effective Hamiltonian, we identify topological phases based on the sign of the coin parameter 
$\theta$. Central to our study is the existence of edge states localized at interfaces between topologically distinct phases, which persist even in inhomogeneous systems.

We further explore how finite-size effects influence topological properties. In small systems, the distinction between topological phases becomes less pronounced, and gap states can persist even when system parameters suggest otherwise. By analyzing the dependence of edge state energies on system parameters, we demonstrate how finite-size effects break degeneracy and lead to non-trivial energy spectra.

Finally, we examine the dynamics of DTQWs on finite lattices with reflecting boundaries. When the boundary coins share the same topological phase as the bulk, ballistic transport ("ping-pong" motion) dominates. In contrast, when the boundary coins are in a different topological phase, the population can become trapped in a bi-localized superposition of edge states, exhibiting coherent oscillations between them. We term this effect Rabi transport, as it involves the coherent transfer of population between lattice ends without populating the central region. This phenomenon arises from the coupling of degenerate edge states, with the oscillation period exponentially dependent on the lattice size.

In the case of homogeneous bulk, we provide a fully analytical solution. Moreover, we demonstrate with the use of numerical simulations the robustness of Rabi transport against random variations in coin parameters, provided the bulk remains in a distinct topological phase from the boundaries. This underscores the topological protection of the observed phenomena.

Our study provides a comprehensive analysis of topological effects in DTQWs, emphasizing the roles of edge states, finite-size effects, and the interplay between topology and quantum transport. These findings have significant implications for quantum simulation and the design of robust quantum systems with topologically protected states.

\section{Appendix - Quantum walk on finite lattice}

In this appendix, we consider a quantum walk on a finite, one-dimensional lattice. We aim to find the eigenvalues and eigenvectors of such a model.

\subsection{Model with an odd number of vertices and real coins}

Let us start with a wire with an odd number ($2L+3$) of vertices and real coins
\begin{equation}\label{coin}
    C_x = \begin{pmatrix}\cos \theta_x& \sin \theta_x\\-\sin \theta_x & \cos \theta_x\end{pmatrix}=e^{i \theta_x \sigma_y}\in \mathcal{SU}(2)
\end{equation}
with $\theta_x=\theta>0$ for $-L \leq x \leq L$ and $\theta_{\pm(L+1)}=-\frac{\pi}{2}$.
We denote the stationary (although not normalized) states with energy $\omega$ by
\begin{equation}
    \ket{\Psi_{\omega}} = \sum_x \ket{x}\otimes\begin{pmatrix} a_x\\b_x\end{pmatrix}.
\end{equation}
Note that it must be $a_{-L-1}= b_{L+1}=0$.

Let us highlight now the symmetries of the system that are crucial to solving our problem.

First, the system possesses so-called particle-hole symmetry (PHS) which means that the unitary evolution operator satisfies
\begin{equation}
  \Omega U \Omega^{-1}=U,
\end{equation}
where $\Omega$ denotes complex conjugation in position and coin basis
\begin{equation}
    |x\rangle \otimes \begin{pmatrix}1\\0\end{pmatrix} ,|x\rangle \otimes \begin{pmatrix}0\\1\end{pmatrix}.
\end{equation}
It follows that the Hamiltonian fulfills
\begin{equation}
  \Omega H \Omega^{-1}=-H.
\end{equation}
There is also sublattice symmetry
\begin{equation}
   \Lambda U\Lambda^{-1}= -U
\end{equation}
with
\begin{equation}
    \Lambda = \sum_x (-1)^{x}\ket{x}\bra{x} \otimes I.
\end{equation}
Another important symmetry is parity symmetry.
For $C_x=C_{-x}$ (which is satisfied in the considered model), one has
\begin{equation}
  \Pi U \Pi^{-1}=U
\end{equation}
and
\begin{equation}
  \Pi H \Pi^{-1}=H
\end{equation}
where unitary
\begin{equation}\label{parity}
   \Pi = \sum_x \ket{-x}\bra{x} \otimes \sigma_y.
\end{equation}

We will also use a unitary operator
\begin{equation}
   \Gamma =S(I \otimes \sigma_x).
\end{equation}
which gives chiral symmetry
\begin{equation}
  \Gamma U \Gamma^{-1}=U^{-1}
\end{equation}
and
\begin{equation}
  \Gamma H \Gamma^{-1}=-H.
\end{equation}
The above symmetries fulfill $\Omega^2=\Lambda^2=\Pi^2=\Gamma^2=I$. 
With the use of the above symmetries for each $\ket{\Psi_{\omega}}$ one can define three other eigenstates of the walk
\begin{equation}\label{68}
  \ket{\Psi_{-\omega}}=\Omega  \ket{\Psi_{\omega}},
\end{equation}
\begin{equation}\label{69}
  \ket{\Psi_{\omega-\pi}}=\Lambda  \ket{\Psi_{\omega}}
\end{equation}
and
\begin{equation}\label{70}
  \ket{\Psi_{-(\omega-\pi)}}=\Lambda \Omega  \ket{\Psi_{\omega}}.
\end{equation}
\subsection{Toward solution - eigenvector form}
Observe that the chiral partner of $\ket{\Psi_{-\omega}}$ must be proportional to $\ket{\Psi_{\omega}}$ i.e.
\begin{equation}
 \ket{\Psi_{\omega}}=\Omega  \ket{\Psi_{-\omega}}=e^{i \tau}\Gamma \ket{\Psi_{-\omega}}.
\end{equation}
With the proper choice of global phase, one can eliminate phase $\tau$ and obtain the condition
\begin{equation}\label{sym1}
 b_x=a_{x+1}^*.
\end{equation}
Moreover, from parity symmetry we have (or we may have in the case of degeneracy)
\begin{equation}
  \Pi\ket{\Psi_{\omega}}=\mu \ket{\Psi_{\omega}}
\end{equation}
with $\mu=\pm 1$, from which it follows that
\begin{equation}\label{par}
  i a_x=\mu b_{-x}.
\end{equation}
Now condition
\begin{equation}
    U \ket{\Psi_{\omega}}=\lambda \ket{\Psi_{\omega}},
\end{equation}
with eigenvalue $\lambda = e^{-i\omega}$ 
can be formulated in the following way 
\begin{equation}\label{amp}
    a_{L+1}=\lambda b_{L}=\lambda  a_{L+1}^*
\end{equation}
and for $x \geq -L$
\begin{equation}\label{equ}
    c \ a_x+s \ a_{x+1}^* =\lambda \ a_{x+1},
\end{equation}
where $c=\cos \theta$ and $s=\sin \theta$. Let
\begin{equation}
 a_x=r_x+ip_x
\end{equation}
with real $r_x$ and $p_x$ and let $\alpha=\cos \omega$ and $\beta=\sin \omega$.
Then Eq. \ref{equ} can be rewritten in the form
\begin{eqnarray}
\beta \ p_{x+1}=c r_x+(s-\alpha) r_{x+1}\\
-\beta \ r_{x+1}=c p_x-(s+\alpha) p_{x+1}
\end{eqnarray}
It follows that
\begin{equation}\label{px}
p_{x}=\frac{s+ \alpha}{\beta}r_{x}-\frac{c}{\beta}r_{x+1}
\end{equation}
and
\begin{equation}\label{er}
 r_{x+2}-\frac{2\alpha}{c}r_{x+1}+r_x=0.
\end{equation}
Thus, we obtained a recursive equation that can be solved in a standard way by analyzing the quadratic equation
\begin{equation}
X^2-\frac{2\alpha}{c}X+1=0.
\end{equation}
Two solutions to the above equation fulfill $X_1 \cdot X_2=1$ and can be written with the use of the real $k$ as
\begin{equation}
X_{1/2}=e^{\pm k}
\end{equation}
with
\begin{equation}
cosh(k)=\frac{\alpha}{c}
\end{equation}
for $\alpha > c$ or
\begin{equation}
X_{1/2}=e^{\pm i k}
\end{equation}
with
\begin{equation}
cos(k)=\frac{\alpha}{c}
\end{equation}
for $\alpha < c$.
There is also another solution
\begin{equation}
X_{1/2}=1
\end{equation}
for $\alpha = c$.
\subsection{Gap state}
Below, we concentrate on the stationary state with positive energy $\omega$ in the $E=0$ gap.
Let us consider the case $\alpha > c$. The solution of Eq. \ref{er} is of the form (with real $r_+,r_-$)
\begin{equation}
r_x=r_+ \ e^{kx}+r_- \ e^{-kx}.
\end{equation}
From Eq. \ref{px}
\begin{equation}
p_x=\left(\frac{s+\alpha-c e^k}{\beta}\right)r_+ \ e^{kx}+\left(\frac{s+\alpha-c e^{-k}}{\beta}\right)r_- \ e^{-kx}
\end{equation}
thus
\begin{equation}
a_x=a_+ \ e^{kx}+a_- \ e^{-kx}
\end{equation}
with
\begin{equation}\label{eqa}
a_{\pm}=|a_{\pm}| e^{i\chi_{\pm}}=\frac{r_{\pm}}{\beta} \left(\beta+i (s \mp c \ sinh(k))\right).
\end{equation}
Let us define $\chi_0$ and $\phi$ via
\begin{equation}
\chi_{\pm}=\chi_0 \pm \phi.
\end{equation}
One has
\begin{equation}\label{eqa}
a_+ a_-=|a_+| | a_-| e^{2 i \chi_0}=i \ \frac{2 \ s \ r_+ \ r_-}{\beta}.
\end{equation}
It must be $\chi_0 =z \frac{\pi}{4}$,
where $z=sgn(r_+ r_-)$ (note that $\beta >0$ with our assumption of $\omega >0$). On the other hand
\begin{equation}\label{eqa}
a_+ a_-^*=|a_+| | a_-| e^{2 i \phi}=2 r_+ r_- \left(1-i \frac{c \ sinh(k)}{\beta} \right).
\end{equation}
Let us define $\chi$ (without subscript) with the equation (note that $\alpha >c$ guarantees $\beta < s$)
\begin{equation}
s \ \sin \chi=\beta.
\end{equation}
 Then 
\begin{equation}\label{eqa}
 \frac{c \ sinh(k)}{\beta}=\frac{cos \chi}{sin \chi},
\end{equation}
and
\begin{equation}\label{eqa}
a_+ a_-^*=-i \frac{2 r_+ r_-}{sin \chi} \left(cos \chi +i sin \chi \right).
\end{equation}
It follows that
\begin{equation}\label{eqa}
\phi= \frac{\chi}{2}-z\frac{\pi}{4} .
\end{equation}
Now we can write
\begin{equation}\label{ax}
a_x=e^{i z \frac{\pi}{4}}\left(|a_+| \ e^{kx+i\phi}+|a_-| \ e^{-(kx+i\phi)}\right)
\end{equation}
Let us assume that $\ket{\Psi_{\omega}}$ is an eigenstate of the parity operator with eigenvalue $\mu$. It follows from Eqs. \ref{sym1} and \ref{par} that
\begin{equation}\label{sym2}
i a_x= \mu a_{1-x}^*.
\end{equation} 
Equation above with $a_x$ taken from Eq. \ref{ax} gives
\begin{equation}
\left(|a_+|e^k+z \mu|a_-|\right)e^{-(kx+i\phi)}=-\left(z \mu|a_+|+|a_+|e^{-k}\right)e^{(kx+i\phi)}.
\end{equation}
To fulfill it for arbitrary $x$ (and $k \neq0$) it must be
\begin{equation}\label{104}
|a_+|e^k+z \mu|a_-|=0.
\end{equation}
For which solution exists only for $z \mu=-1$.
\\For unnormalized states, we are free to choose
\begin{equation}\label{105}
|a_{\pm}|=e^{\mp k/2}.
\end{equation}
Finally,
\begin{equation}\label{ax1}
a_x=e^{-i \frac{\pi}{4}}\left(e^{k(x-\frac{1}{2})+i\phi}+e^{-(k(x-\frac{1}{2})+i\phi)}\right),
\end{equation}
with $\phi= \frac{\chi}{2}+\frac{\pi}{4}$.
\subsection{Eigenvalue}
From Eq. \ref{amp} we have 
\begin{equation}
a_{L+1}=|a_{L+1}|e^{-i \omega /2}.
\end{equation}
On the other hand from Eqs. \ref{ax1} we have
\begin{equation}
a_{L+1}=|a_{L+1}|e^{-i \pi /4} e^{i \alpha},
\end{equation}
with
\begin{equation}
\tan \alpha=\tan \phi \  \tanh(kZ),
\end{equation}
where $Z=L+\frac{1}{2}$.
Consequently
\begin{equation}\label{tan}
\tan \left( \frac{\omega}{2}-\frac{\pi}{4} \right)=-\tan \left( \frac{\chi}{2}+\frac{\pi}{4} \right) \  \tanh(kZ).
\end{equation}
Let us recall that
\begin{equation}
 \sin \chi=\frac{\sin \omega}{\sin \theta},   
\end{equation}
and
\begin{equation}\label{end}
 \cosh k=\frac{\cos \omega}{\cos \theta},   
\end{equation}
System of Eqs. \ref{tan} - \ref{end} can be numerically solved to give $\omega$ and $k$. 

\subsection{Approximate solution}

\begin{figure}[t]
\includegraphics[width=8cm]{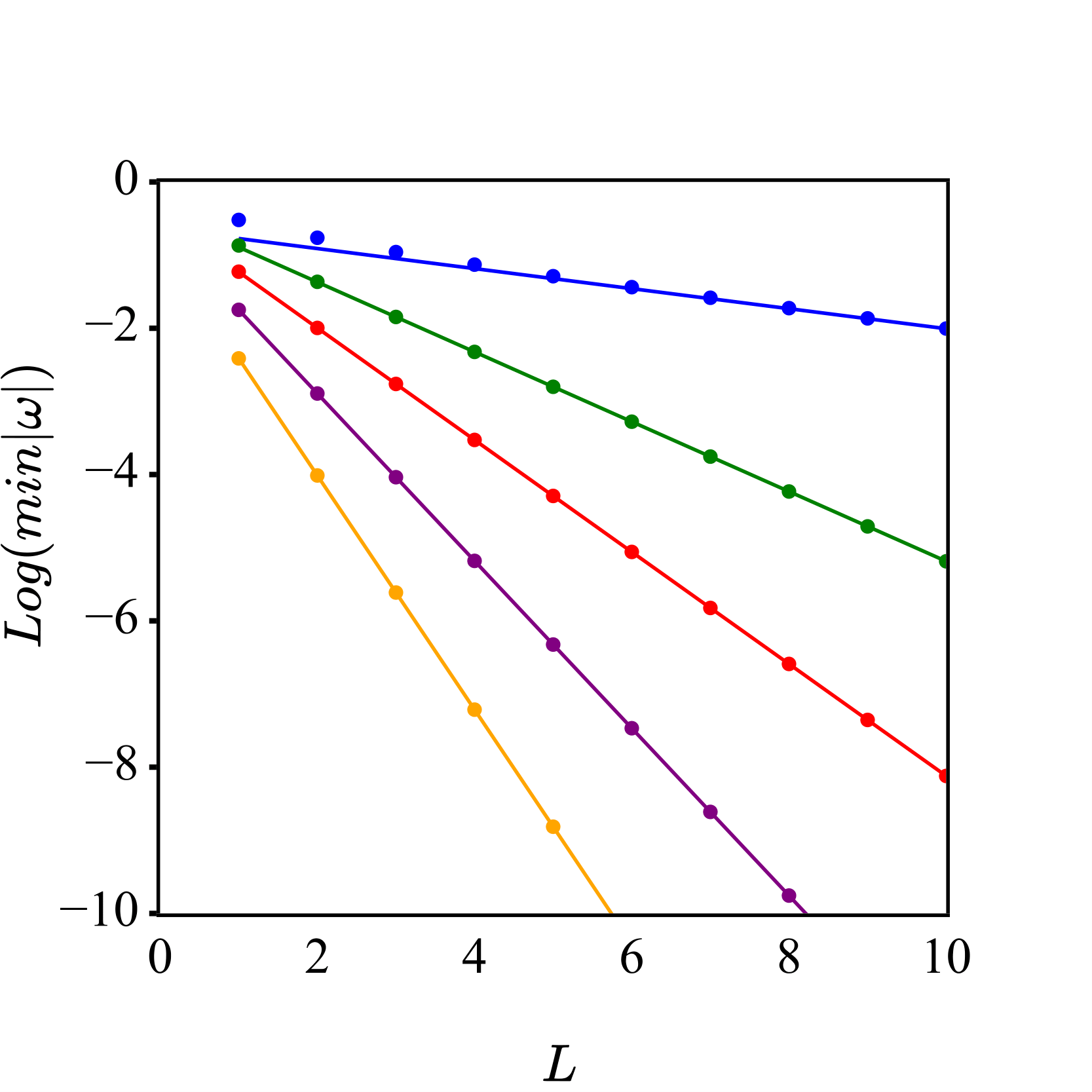}
\caption{Logarithm of the smallest absolute value of the eigenenergy versus $L$ for different $\theta$ (from highest value of the aforementioned logarithm: $\frac{\pi}{20}$ (blue), $\frac{\pi}{6}$ (green), $\frac{\pi}{4}$ (red), $\frac{\pi}{3}$ (purple), $\frac{2\pi}{5}$ (orange). Numerical sollutions of Eqs. \ref{tan} - \ref{end} - dots, aproximate values of Eq. \ref{appr} - lines.}
\label{fig20}
\end{figure}

Approximate solution of the system of Eqs. \ref{tan} - \ref{end}  can be obtained with the assumption that $\omega$ is close to zero. In this case (assuming also that $\omega << \theta$ and $Z>>1$)
\begin{equation}
\tan \gamma \approx -(1-\omega),
\end{equation}
\begin{equation}
\tan \phi \approx 1+\frac{\omega}{\sin \theta},
\end{equation}
\begin{equation}
\tanh(kZ) \approx 1-2 e^{-2 k_0 Z},
\end{equation}
where
\begin{equation}
e^{k_0} \equiv \frac{1+\sin \theta}{\cos \theta}.
\end{equation}
This leads to the formula
\begin{equation}\label{appr}
\omega \approx \omega_0 \equiv \frac{\sin 2 \theta}{(1+\sin \theta)^2}  e^{-2 k_0 L}=\frac{\sin 2 \theta}{(1+\sin \theta)^2}\left(\frac{1+\sin \theta}{\cos \theta} \right)^{-2 L}.
\end{equation}
In the above approximation, amplitudes of stationary states are given by 
\begin{equation}\label{ax2}
a_x=e^{k_0(x-\frac{1}{2})+i\phi_0}-ie^{-k_0(x-\frac{1}{2})+i\phi_0},
\end{equation}
\begin{equation}
b_x=e^{k_0(x+\frac{1}{2})-i\phi_0}+ie^{-k_0(x+\frac{1}{2})-i\phi_0},
\end{equation}
where
\begin{equation}
\phi_0 \equiv \frac{\cos \theta}{(1+\sin \theta)^2}  e^{-2 k_0 L}.
\end{equation}
In Fig. \ref{fig20} we present 
\subsection{Model with even number of vertices}
In the case of even number of vertices, we choose $\theta_x=\theta>0$ for $-L+1 \leq x \leq L$ and $\theta_{-L}=\theta_{L+1}=-\frac{\pi}{2}$. To obtain stationary solutions we can follow previous (odd case) arguments with slight modifications. The parity operator (Eq. \ref{parity}) must take the form
\begin{equation}
   \Pi = \sum_x \ket{1-x}\bra{x} \otimes \sigma_y.
\end{equation}
and it follows that
\begin{equation}
  a_x=(\pm 1)(-i )b_{1-x}.
\end{equation}
In consequence Eqs. \ref{sym1} and \ref{sym2} must be replaced by
\begin{equation}
b_{1-x}=i a_x
\end{equation} 
and 
\begin{equation}
a^*_{2-x}=i a_x
\end{equation}
respectively. This leads to the condition
\begin{equation}
|a_+|e^{2k}-|a_-|=0.
\end{equation}
which can be fulfilled by
\begin{equation}
|a_{\pm}|=e^{\mp k}.
\end{equation}

\subsection{Unified (odd and even case) notation}
In this section, we provide unified notation valid for both odd and even cases, using $\zeta=1$ for the odd and $\zeta=2$ even cases.
\begin{equation}
    \ket{\Psi_{\omega}} = \sum_x \ket{x}\otimes\begin{pmatrix} a_x\\a^*_{x+1}\end{pmatrix},
\end{equation}
with
\begin{equation}\label{ax11}
a_x=e^{-i \frac{\pi}{4}}\left(e^{k(x-\frac{\zeta}{2})+i\phi}+e^{-(k(x-\frac{\zeta}{2})+i\phi)}\right).
\end{equation}
Unified version of Eq. \ref{tan} is
\begin{equation}
\tan \gamma=-\tan \phi \  \tanh(kZ_{\zeta}),
\end{equation}
where 
\begin{equation}
Z_{\zeta}=L+1-\frac{\zeta}{2},
\end{equation}
Unified version of approximate energy (Eq. \ref{appr}) is
\begin{equation}
\omega_0 = 2 \  \tan \theta \left( \frac{1+sin \theta}{cos \theta} \right)^{\zeta-3}  e^{-2 k_0 L}.
\end{equation}

\subsection{Model with general SU(2) coin}
Let us abandon the assumption that a coin must be real. We will consider coin of the form
\begin{equation}\label{coin}
    C_x = \begin{pmatrix}e^{i \zeta}\cos\theta_x & e^{i (\zeta+\sigma)}\sin\theta_x\\-e^{-i (\zeta+\sigma)}\sin\theta_x & e^{-i \zeta}\cos\theta_x\end{pmatrix},
\end{equation}
where phases $\zeta$ and $\sigma$ are assumed to be $x$-independent. Within this assumption, the above coin is a general SU(2) coin and, moreover, is equal to any U(2) coin up to phase. Let us use the following form of the states
\begin{equation}
    \ket{\Psi} = \sum_x e^{i \zeta x} \ket{x}\otimes\begin{pmatrix} a_x\\ e^{- i \sigma} b_x\end{pmatrix}.
\end{equation}
The equations for the amplitudes $a_x$ and $b_x$ turn out to be the same as in the real coin case ($\zeta= \sigma =0$)
\begin{equation}
    a_{x,t+1} =\cos \theta_{x-1} a_{x-1,t}+\sin \theta_{x-1} b_{x-1,t}
\end{equation}
\begin{equation}
    b_{x,t+1} =- \sin \theta_{x+1} a_{x+1,t}+\cos \theta_{x+1} b_{x+1,t}.
\end{equation}
It immediately follows that the eigenstate for the general coin is given by
\begin{equation}
    \ket{\Psi_{\omega}} = \sum_x e^{i \zeta x} \ket{x}\otimes\begin{pmatrix} a_x\\ e^{- i \sigma} a^*_{x+1}\end{pmatrix},
\end{equation}
with $a_x$ given by Eqs. \ref{ax11}.
\subsection{Remaining gap states}
Now we can use Eqs. \ref{68} - \ref{70} to obtain remaining three gap states. However, in the case of general coin PHS operator must be generalized to the form
\begin{equation}
   \Omega'  =\left( \left( \sum_x e^{2i \zeta x } \ket{x} \bra{x} \right)\otimes \begin{pmatrix}1&0\\0&e^{-2i\sigma}\end{pmatrix} \right) \Omega,
\end{equation}
We obtain
\begin{equation}
    \ket{\Psi_{-\omega}} =\Omega'  \ket{\Psi_{\omega}}=\sum_x e^{i \zeta x} \ket{x}\otimes\begin{pmatrix} a^*_x\\ e^{- i \sigma} a_{x+1}\end{pmatrix},
\end{equation}
\begin{equation}
    \ket{\Psi_{\omega-\pi}} =\Lambda  \ket{\Psi_{\omega}}=\sum_x (-1)^x e^{i \zeta x} \ket{x}\otimes\begin{pmatrix} a_x\\ e^{- i \sigma} a^*_{x+1}\end{pmatrix}
\end{equation}
and
\begin{equation}
    \ket{\Psi_{-(\omega-\pi)}} =\Lambda \Omega \ket{\Psi_{\omega}}=\sum_x (-1)^x e^{i \zeta x} \ket{x}\otimes\begin{pmatrix} a^*_x\\ 
    e^{- i \sigma} a_{x+1}\end{pmatrix},
\end{equation}
\subsection{Band states}
Let us now return to the simplest mode with an odd number of vertices and real coins but this time let us 
consider the case $\alpha < c$. The solution of Eq. \ref{er} is now of the form 
\begin{equation}
r_x=\sum_{\pm} r_{\pm} \ e^{{\pm}ikx}
\end{equation}
with $r_-=r_+^*$.
From Eq. \ref{px}
\begin{equation}
p_x=\sum_{\pm}\left(\frac{s+\alpha-c e^{{\pm}ik}}{\beta}\right)r_{\pm} \ e^{{\pm}ikx}
\end{equation}
and
\begin{equation}
a_x=\sum_{\pm}a_{\pm} \ e^{{\pm}ikx}
\end{equation}
with
\begin{equation}\label{eqa}
a_{\pm}=|a_{\pm}| e^{i\chi_{\pm}}=\frac{r_{\pm}}{\beta}\left(\beta \pm c \ sin(k) + i s \right).
\end{equation}
Let us use
\begin{equation}
\chi_{\pm}=\chi_0 \pm (\phi-\frac{k}{2}).
\end{equation}
The utility of putting $\frac{k}{2}$ in the above equation will appear soon.
Note that
\begin{equation}
a_+ a_-=|a_+| | a_-| e^{2 i \chi_0}=i\frac{2 s | r_+| | r_-|}{\beta}.
\end{equation}
So it must be $\chi_0 = \frac{\pi}{4}$.
On the other hand
\begin{equation}\
|a_{\pm}|^2=\frac{|r_{\pm}|^2}{\beta^2} \left((\beta \pm c \ sin(k))^2+s^2 \right).
\end{equation}
Let us (now in a different way) define $\chi$ (without subscript) with equation (note that $\alpha <c$ guaranteed that $\beta > s$)
\begin{equation}\label{sinchi}
\beta \ \sin \chi=s.
\end{equation}
 Then 
\begin{equation}\label{eqa}
|a_{\pm}|^2=2 |r_{\pm}|^2 (1 \pm cos \chi).
\end{equation}
Without normalization condition, we are free to choose $2 |r_{\pm}|^2=1$ and obtain
\begin{equation}\label{eqa}
|a_{\pm}|=\sqrt{ (1 \pm cos \chi)}.
\end{equation}
 Now we can write
\begin{equation}\label{aka1}
a_x=e^{i \frac{\pi}{4}}\sum_{\pm} \sqrt{ (1 {\pm} cos \chi)} \ e^{{\pm}i\left( k(x-\frac{1}{2})+\phi \right)}.
\end{equation}
Let us assume that $\ket{\Psi_{\omega}}$ is the eigenstate of the parity operator with eigenvalue $\mu$. Using Eqs. \ref{sym2} and \ref{aka1} we obtain the condition
\begin{equation}
- \mu e^{i (\phi-\frac{k}{2})}=e^{-i (\phi-\frac{k}{2})} e^{-ik},
\end{equation}
so
\begin{equation}
\phi=(1+\mu)\frac{\pi}{4}.
\end{equation}

From Eq. \ref{amp} we have again
\begin{equation}
a_{L+1}=|a_{L+1}|e^{-i \omega /2}.
\end{equation}
On the other hand, from Eqs. \ref{aka1} we have
\begin{equation}
a_{L+1}=|a_{L+1}|e^{i \pi /4} e^{i \psi},
\end{equation}
with
\begin{equation}
\tan \psi=\tan (kZ+\phi) \  \frac{cos \chi}{1+sin \chi},
\end{equation}
where $Z=L+\frac{1}{2}$.
In a consequence (with the use of Eq. \ref{sinchi})
\begin{equation}\label{tan1}
\tan \left( \frac{\omega}{2}+\frac{\pi}{4} \right)\  \sqrt{\frac{\sin \omega +\sin \theta}{\sin \omega -\sin \theta}}=-\tan \left( kZ+(1+\mu)\frac{\pi}{4} \right) .
\end{equation}
Let us recall that
\begin{equation}\label{end1}
 \cos k=\frac{\cos \omega}{\cos \theta}.  
\end{equation}
System of Eqs. \ref{tan1} - \ref{end1} can be numerically solved to give $\omega$ and $k$. 

Similarly to the case of gap states above, solutions can be generalized to an even number of vertices and a general $\mathcal{U}(2)$ coin.

Finally, all obtained states can be normalized in a standard way.
\section{Data Availability Statement}
The data that support the findings of this article are openly available at \cite{data}
\section{Acknowledgements}
We want to express our gratitude to Adam Sajna for insightful discussions. This research is supported by the Polish National Science Centre (NCN) under the Maestro Grant no. DEC-2019/34/A/ST2/00081.

\end{document}